\documentclass[proceedings]{JHEP3} 

\usepackage{epsfig,multicol}            

\newbox\mybox
\newcommand{\ttbs}{\char'134}           
\newcommand\fverb{\setbox\mybox=\hbox\bgroup\verb}
\newcommand\fverbdo{\egroup\medskip\noindent\fbox{\unhbox\mybox}\ }
\newcommand\fverbit{\egroup\item[\fbox{\unhbox\mybox}]}

\conference{International Workshop on Astroparticle and High
Energy Physics}

\title{Neutrino induced showering from the Earth}

\author{\speaker{Daniele Fargion}\\
    Physics Department and INFN, University "La Sapienza", P.le A. Moro,2, 00185 ROME,
    ITALY\\
    E-mail: \email{daniele.fargion@roma1.infn.it}}

\abstract{Ultra High Energy, UHE, Neutrino Astronomy should be
soon tested looking toward the Earth. At  present High Energy
Neutrino Astronomy is searched by AMANDA, ANTARES underground
detectors looking for its consequent unique muons secondary
track. We suggest a higher energy Tau Neutrino Astronomy based on
Horizontal and Upward Tau Air-Showers escaping from the Earth.
These Tau air-showers greatly amplifies the single tau track by an
abundant secondary tail ( billions of electron pairs, gamma and
tens of millions muon bundles) spread in huge areas (kilometer
size) easily observable (even partially) from high mountains,
balloon or satellite array detectors. Possible early evidence of
such a New Neutrino UPTAUs or HORTAUs (Upward or Horizontal Tau
Air-Showers) Astronomy may be already found in rare BATSE gamma
records of brief up-going gamma showers named Terrestrial Gamma
Flashes (TGF). The TGF features, energy and arrival clustering
are well tuned to upward tau air-showers. Future confirmation of
the Neutrino Tau Astronomy must be found in detectors as EUSO,
seeking for Upward-Horizontal air-shower events: indeed even
minimal, guaranteed, GZK neutrino fluxes may be observed if EUSO
threshold reaches $10^{19}$ eV or lower energies by enlarging its
telescope size.}

\begin{document}

\section{ From Cosmic Rays to Ultra High Energy Tau Neutrinos Astronomy}
Nature, for unknown reasons or because inflation in early
Universe, did not provide (much or at all)
 magnetic monopoles in our Universe.
Their absence is proved by  the existence of coherent large scale
structure of magnetic fields near  planets, stars and galactic
spaces (the so called Parker limit). Cosmic Rays, because of their
main nucleon and nuclei nature, are  charged and consequently
they are easily bent by solar, galactic and  extra-galactic
magnetic fields. Therefore   Cosmic Rays are reaching our
terrestrial atmosphere after a long random walk by Lorents forces,
smeared and nearly isotropic with no astronomical legacy.

 Astronomy, on the contrary, is based on the
direct, undeflected radiation from astronomical source to the
observer. The neutral messengers which we did exploit in
Astronomy during last three centuries were mainly photons in the
optical range. Since Maxwell discovers and along last century the
whole electromagnetic spectrum was opened to Astronomy. Radio,
Infrared , X and gamma astronomy made vivid in new colors hidden
sides  of our sky. Photons are not the unique neutral particle.
Indeed in the last $30$ years also neutrino astronomy arose and
proved fruit-fully in discovering solar and supernova neutrino
astronomy at low (MeV) energies. Also at the extreme energy edges,
Ultra High Energy Cosmic Ray (UHECR) may offer a New Particle
Astronomy because, even they are charged, their extreme rigidity
makes them almost un-bent by galactic and extra-galactic magnetic
fields. Therefore UHECR, even charged, fly nearly straight from
their birth-place to us offering a New Particle Astronomy. These
UHECR ($E _{UHECR} \geq 10^{19} eV$) are numerous and they are
already reaching hundreds of data records. Moreover this UHECR
astronomy is bounded (by primordial photon drag, the well known
Greisen, Zatsepin, Kuzmin (GZK) cut-off
\cite{Greisen:1966jv},\cite{Zatsepin:1966jv}. ) in a very narrow
(almost local) Universe a few tens of Mpc wide. Surprisingly we
did't find yet in present UHECR arrival maps  any corresponding
nearby known galactic  or  super-galactic imprint. The observed
UHECR are coming from everywhere. This  isotropy call naturally
for a cosmic link, whose distances are well above the narrow GZK
radius cut-off. Moreover observed UHECR clustering in groups
strongly suggest compact sources (as AGN or BL Lacs beaming Jets)
respect to any homogeneous and isotropic halo made by  primordial
topological defects.  Recent evidence for a dozen or more BL Lacs
sources correlation (some at medium redshift $z \simeq 0.3 \gg
z_{GZK}$) with  UHECR clustered events are giving support (with
their over all isotropy) to a cosmic
 origin for UHECR sources \cite{Kalashev:2002kx}.
This is the so-called GZK paradox that may find a solution by
neutrino with a light mass, in the so called Z-Burst model.
Indeed  UHE Relic neutrinos with a light mass may play a role as
calorimeter for UHE neutrinos from cosmic distances at ZeV
energies. Their scattering may solve the GZK paradox,
\cite{Fargion Salis 1997}, \cite{Fargion Mele Salis 1999},
\cite{Weiler 1999}, \cite{Yoshida  et all 1998}, in particular in
a narrow neutrino mass windows \cite{Fargion et all. 2001b},
\cite{Fodor Katz Ringwald 2002}; in this scenario UHE $\nu$
Astronomy is not just a consequence but itself the cause of UHECR
signals. There are other additional solution to GZK paradox based
on extreme extragalactic magnetic fields, new hadronic physics or
hypothetical  decay of primordial topological relics; we shall
not address here to these models for lack of space; anyway large
neutrino secondaries or primaries fluxes must also arise
\cite{Kalashev:2002kx}in most models. Indeed the same UHECR
constrained by GZK cut off must produce abundantly by pion decays
the so called GZK (or cosmogenic) neutrino secondaries at a
fluence at least comparable with most penetrating  UHECR.
Therefore there must be a minimal underlying neutrino astronomy
nearly as large as observed GZK cosmic ray level  (at energy
$E\leq 10^{19} eV$). In order to solve the GZK puzzles  it may be
necessary to  trace their linked UHE neutrinos either primary (at
much higher fluxes)  by Z-Showering or Z-Burst model or, for GZK
neutrinos at least as secondaries of the same UHECR. To test this
idea we need to open an independent UHE neutrino Astronomy on
Earth. Different muon track detectors, cubic $km^3$ in
underground may be reveal them at PeVs energies. Nevertheless UHE
neutrino tau may interact on Mountains or better on Earth crust,
at a huge concrete calorimeter, leading to Tau Air-Showers,
probing easily both (secondary and/or primary)  UHE neutrino
astronomy  at GZK energies. Present article review  the problems
and the experimental prospects on Mountains, planes detectors
quota;  we also reconsidered events measured by satellites,
 like  past BATSE, present gamma satellite INTEGRAL,  future EUSO  experiments. The recent
signals  in BATSE terrestrial gamma flashes maybe indeed the
first evidence for these new UHE neutrino Astronomy upcoming from
Earth crust at PeV-EeV energies, leading to UPTAUs and HORTAUs
showers.  Indeed in last Fig.\ref{fig:fig17} we
 summirized the two consequent Flux signal derived by the TGF event rate data in
BATSE (1991-2000) experiment, normalized for the estimated BATSE
thresholds. These preliminary result may be a useful reference
estimate for PeV-EeV neutrino fluxes and apparently they are well
consistent with Z-Showering model for a relic mass within the
expected values ($m \simeq 0.04-0.4 eV$).
 Let us remind that the restrictions on any
astronomy are related to a the messenger interactions  with the
surrounding  medium on the way to the observer. While Cosmic Rays
astronomy is severely blurred by random terrestrial, solar,
galactic and extragalactic  magnetic lenses, the highest $\gamma$
ray astronomy (above tens TeV) became more or less  blind because
of photon-photon opacity (due to electron pair production) at
different energy windows. Indeed the Infrared- TeV opacity as
well as a more severe Black Body Radiation, BBR,( at $2.75
K$)-PeV cut-off are bounding the TeV -PeV $\gamma$ ray astronomy
in  very nearby cosmic ( or even galactic) volumes. Therefore
rarest TeV gamma signals are at present the most extreme  trace
of High Energy Astronomy. We observe copious cosmic rays at
higher ($\gg 10^{15}$eV) energies almost isotropically spread by
galactic  and cosmic magnetic fields in the sky.
 Let us remind,
among  the $\gamma$ TeV discoveries, the signals of power-full
Jets blazing to us from Galactic (Micro-Quasars) or extragalactic
edges (BL Lacs). At PeV energies astrophysical $\gamma$ cosmic
rays should also be presented, but, excluding a very rare and
elusive Cyg$X3$ event, they have not being up date observed; only
upper bounds are known at PeV energies.  The missing $\gamma$ PeV
astronomy, as we mentioned, are very probably absorbed because of
their own photon interactions (electron pairs creation) at the
source environment and/or along the photon propagation into the
cosmic Black Body Radiation (BBR) and/or into other diffused
background radiation. Unfortunately PeV charged cosmic rays,
easily bend and bounded in a random walk by Galactic magnetic
fields, loose their original directionality and their
astronomical relevance; their tangled trajectory resident time in
the Galaxy is much longer ($\geq 10^{3}$ - $10^{5}$) than any
linear neutral trajectory, as in the case of gamma rays, making
the charged cosmic rays  more probable to be observed by nearly a
comparable length ratio. However astrophysical UHE neutrino
signals in the wide range $10^{13}$eV-$10^{19}$eV (or even higher
GZK energies) are unaffected by any radiation cosmic opacity and
may easily open a very new exciting window toward Highest Energy
sources. Being weakly interacting the neutrinos are an ideal
microscope to deeply observe  in their accelerator (Jet,SN,GRB,
Mini Black Hole) cores they do not experience any strong self
opacity as the case of photon. Other astrophysical $\nu$ sources
at lower energies ($10^{8}$ eV - $10^{12}$ eV) should also be
present, at least at EGRET fluence level, but their signals are
very weak and probably drowned by the dominant diffused
atmospheric $\nu$, secondaries of muon secondaries, produced as
pion decays by the same charged (and smeared) UHE cosmic rays
(while hitting terrestrial atmosphere): the so called diffused
atmospheric neutrinos.
 Indeed a modulation of the
atmospheric neutrinos signal has inferred the first conclusive
evidence for a neutrino mass and for a neutrino flavor mixing. At
lowest (MeVs) $\nu$ energy windows, the abundant and steady solar
neutrino flux, (as well as the prompt, but rare, neutrino burst
from a nearby Super-Novae (SN 1987A)), has been, in last twenty
years, successfully explored, giving support to neutrino flavour
mixing and to the neutrino mass reality. More recent additional
probes of solar neutrino flavour mixing and reactor neutrino
disappearance are giving a robust ground to the neutrino mass
existence, at least at minimal $\sim 0.05$ eV level.

\section{High Energy Tau Showering on top Earth Atmosphere}
\begin{figure}
\centering\includegraphics[width= 9cm]{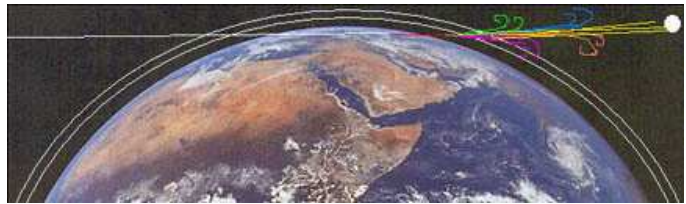}
\caption {Horizontal Upward Tau Air-Shower (HORTAUS) originated
by UHE neutrino skimming the Earth: fan-like jets due to
geo-magnetic bending  shower at high quota ($\sim 23-40 km$):
they may be pointing to an orbital satellite detector . The
Shower tails may be also observable by EUSO just above it.}
\label{fig:fig7}
\end{figure}

\begin{figure}
\vspace{-0.2cm}
\centering\includegraphics[width=9cm]{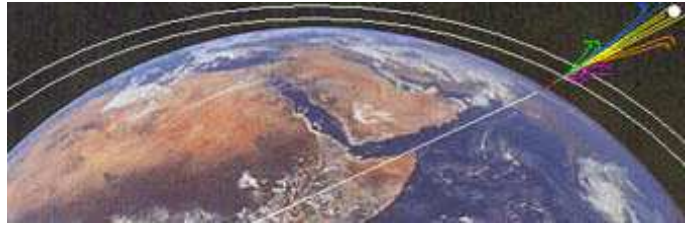}
\caption {A very schematic Upward Tau Air-Shower (UPTAUs)  and
its open fan-like jets due to geo-magnetic bending at high quota
($\sim 20-30 km$). The gamma Shower may be pointing to an orbital
detector. Its
 vertical Shower tail may be spread by
geo-magnetic field into a thin eight-shape beam observable  by
EUSO  as a small blazing oval (few dot-pixels) aligned orthogonal
to the local magnetic field .} \label{fig:fig8}
 \vspace{0.9cm}
\end{figure}

While longest ${\mu}$ tracks in $km^3$ underground detector have
been, in last three decades, the main searched UHE neutrino
signal, Tau Air-showers by UHE neutrinos generated in Mountain
Chains or within Earth skin crust, see Fig.\ref{fig:fig7} up to
GZK ($>10^{19}$ eV) or upward Fig.\ref{fig:fig8} within PeV-EeV
energies, have been recently proved to be a new powerful amplifier
in Neutrino Astronomy \cite{Fargion et all 1999} \cite{Fargion
2000-2002} \cite{Bertou et all 2002} \cite{Hou Huang 2002}
\cite{Feng et al 2002}. This new Neutrino $\tau$ detector will be
(at least) complementary to present and future, lower energy,
$\nu$ underground  $km^3$ telescope projects (from AMANDA, Baikal,
ANTARES, NESTOR, NEMO, IceCube). In particular Horizontal Tau Air
shower may be naturally originated by UHE $\nu_{\tau}$ at GZK
energies crossing the thin Earth Crust at the Horizon showering
far and high in the atmosphere \cite{Fargion 2000-2002}
\cite{Fargion2001a} \cite{Fargion2001b} \cite{Bertou et all 2002}
\cite{Feng et al 2002}. UHE $\nu_{\tau}$ are abundantly produced
by flavour oscillation and mixing from muon (or electron)
neutrinos, because of the large galactic and cosmic distances
respect to the neutrino  oscillation ones (for already known
neutrino mass splitting). Therefore EUSO may observe many of the
above behaviours and it may constrains among models and fluxes
and it may also answer open standing questions. We will briefly
enlist, in this  presentation, the main different signatures and
rates of UHECR versus UHE $\nu$ shower observable by EUSO at 10\%
duty cycle time within a 3 year record period, offering a more
accurate estimate of their signals.
\begin{figure}
\centering
\includegraphics[width=9cm]{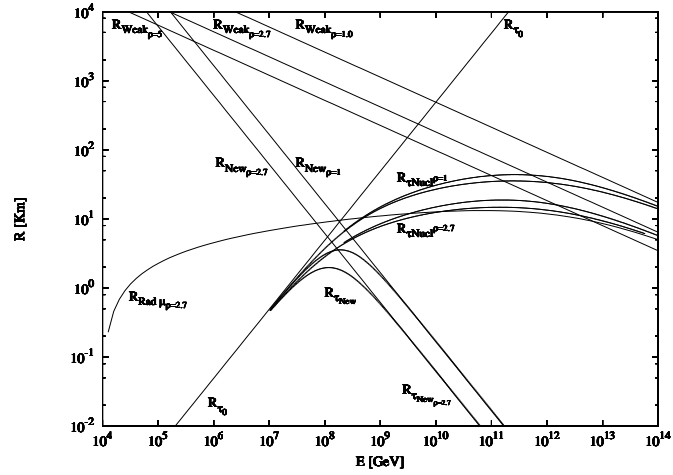}
\caption {Lepton $\tau$ (and $\mu$) Interaction Lengths for
different matter densities: $R_{\tau_{o}}$ is the free $\tau$
length,$R_{\tau_{New}}$ is the New Physics TeV Gravity interaction
range at corresponding densities,$R_{\tau_{Nucl}\cdot{\rho}}$ , is
the combined $\tau$ Ranges keeping care of all known interactions
and lifetime and mainly the photo-nuclear interaction. There are
two slightly different split curves (for each density) by two
comparable approximations in the interaction laws. Note also the
neutrino interaction lenghts above lines $R_{Weak{\rho}}=
L_{\nu}$ due to the electro-weak interactions at corresponding
densities.} \label{fig:fig9}
\end{figure}
\begin{figure}
\centering \vspace{0.5cm}
\includegraphics[width=10cm]{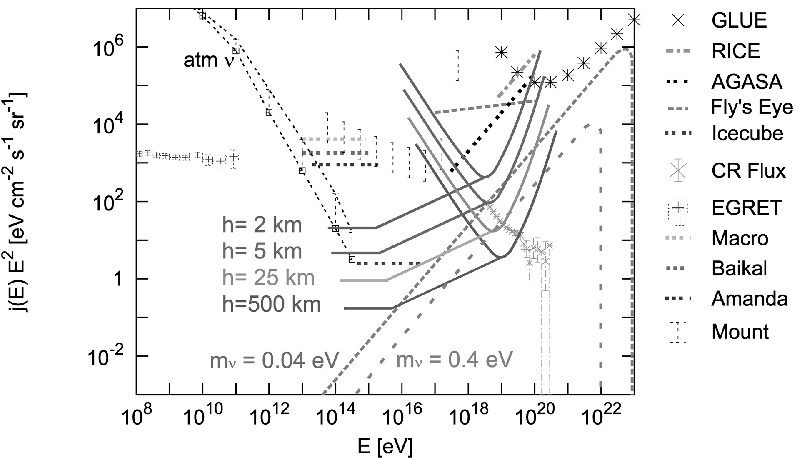}
\caption {UPTAUs (lower bound on the center) and HORTAUs (right
parabolic  curves)  sensibility at different observer heights h
($2,5,25,500 km $) looking at horizons toward Earth seeking upward
Tau Air-Showers adapted over a present neutrino flux estimate in
Z-Shower model scenario for light ($0.4-0.04$ eV) neutrino masses
$m_{\nu}$; two corresponding density contrast for relic light
neutrino masses has been assumed; the lower parabolic bound
thresholds are at different operation height, in Horizontal
(Crown) Detector facing toward most distant horizons edge; these
limits are fine tuned (as discussed in the text) in order to
receive Tau in flight and its
 Shower in the vicinity of the detector; we are assuming a
duration of data records of a decade comparable to the BATSE
record data . The parabolic bounds on the EeV energy range in the
right sides are nearly un-screened by the Earth opacity while the
corresponding UPTAUs bounds  in the center below suffer both of
Earth opacity as well as of a consequent shorter Tau interaction
lenght in Earth Crust, that has been taken into account. }
\label{fig:fig10}
\end{figure}

Let us first consider the last kind of Upward $\tau$ signals due
to their interaction in Air or in Earth Crust
(UPTAUs)Fig.\ref{fig:fig9}. The Earth opacity will filter mainly
$10^{14}\div{10^{16}}$eV upward events \cite{Gandhi et al 1998}
\cite{Halzen1998} \cite{Becattini Bottai 2001} \cite{Dutta et
al.2001} \cite{Fargion 2000-2002}; therefore only the direct
$\nu$ shower in air or the UPTAUs around $3$ PeVs will be able to
flash toward EUSO in a narrow beam ($2.5 \cdot 10^{-5}$ solid
angle) jet blazing apparently at $10^{19}\div{10^{20}}$eV energy.
The shower will be opened in a fan like shape and it will emerge
from the Earth atmosphere spread as a triplet or multi-dot signal
aligned orthogonal to local terrestrial magnetic field. This rare
multi-dot $polarization$ of the outcoming shower will define a
characteristic  signature easily to be revealed. Its opening will
be orthogonal to the magnetic field lines at that particular
geological area. However the effective observed air mass by EUSO
is not $\ 10\%$ (because duty cycle) of the inspected air volume
$\sim 150 km^3$, but because of the narrow blazing shower cone it
corresponds to only to $3.72\cdot 10^{-3}$ $km^3$. The target
volume  increases for upward neutrino Tau interacting vertically
in Earth Crust in last matter layer (either rock or water), see
Fig.\ref{fig:fig9}, making upward relativistic $\simeq 3 PeVs$
$\tau$ whose decay in air born finally an UPTAUs; in this case
the effective target mass is (for water or rock) respectively
$5.5\cdot 10^{-2}$$km^3$ or $1.5 \cdot10^{-1}$ $km^3$.
The characteristic neutrino interaction are partially summirized
in figure above. The consequent $\tau$ and $\mu$ interactions
lenght are also displayed. These lenghts and consequent volume
are not extreme.  The UPTAUs signal is nearly $15$ times larger
than the Air-Induced Upward  $\nu$ Shower hitting Air. The much
wider acceptance of BATSE respect EUSO and the consequent better
threshold (in BATSE) is due to the wider angle view of the gamma
detector, the absence of any suppression factor as in EUSO duty
cycle, as well as the $10$ (for BATSE) over $3$ (for EUSO) years
assumed of record life-time. Any minimal neutrino  fluence
$\Phi_{\nu_{\tau}}$ of PeVs energetic neutrino:
 $ \Phi_{\nu_{\tau}}\geq 10^2 eV cm^{-2} s^{-1}$ might be detectable by EUSO.
\section{Air Induced UHE $\nu$ Shower}
UHE $\nu$ may hit an air nuclei and shower vertically or
horizontally or more rarely nearly up-ward: its trace maybe
observable by EUSO preferentially in inclined or horizontal
case.  Indeed  Vertical Down-ward  ($\theta \leq 60^o$) neutrino
induced Air Shower  occur mainly at lowest quota and they will
only partially shower their UHE $\nu$ energy because of the small
slant depth ($\leq 10^3 g cm^{-2}$) in most vertical down-ward
UHE $\nu$ shower. Therefore the observed  EUSO air mass ($1500
km^3$, corresponding to a $\sim 150$ $km^3$ for $10\%$ EUSO
record time) is only ideally the UHE neutrino calorimeter.
Indeed  inclined $\sim{\theta\geq 60^o }$) and horizontal
Air-Showers ($\sim{\theta\geq 83^o }$) (induced by GZK UHE
neutrino) may reach their maximum output  and their event maybe
observed ; therefore only a small fraction ($\sim 30\%$
corresponding to $\sim 50$ $km^3$ mass-water volume for EUSO
observation) of vertical downward UHE neutrino may be seen by
EUSO. This signal may be somehow hidden or masked by the more
common down-ward UHECR showers.  The key reading  signature will
be the shower height origination: $(\geq 40 km)$ for most
downward-horizontal UHECR,$(\leq 10 km)$ for most
inclined-horizontal Air UHE $\nu$ Induced Shower. A corresponding
smaller fraction ($\sim 7.45\%$) of totally Horizontal UHE
neutrino Air shower, orphan of their final Cherenkov flash, in
competition with the horizontal UHECR, may be also clearly
observed: their observable mass is only $V_{Air-\nu-Hor}$ $\sim
11.1$ $km^3$ for EUSO observation duty-cycle. See
Fig.\ref{fig:fig12}.  A more  rare, but spectacular, double
$\nu_{\tau}$-$\tau$ bang in Air (comparable in principle to the
PeVs expected  "double bang" in water \cite{Learned Pakvasa
1995}) may be exciting, but very
difficult to be observed. \\
The EUSO effective calorimeter mass for such Horizontal event is
only $10\%$ of the UHE $\nu$ Horizontal ones (($\sim 1.1$ $
km^3$)); therefore its event rate is already almost excluded
needing a too high neutrino fluxes (see \cite{Fargion 2002e});
indeed it should be also noted that the EUSO energy threshold
($\geq 3\cdot 10^{19}$eV) imply such a very large ${\tau}$
Lorents boost distance; such large ${\tau}$ track exceed (by more
than a factor three) the EUSO disk Area diameter ($\sim 450$km);
 therefore the expected Double Bang Air-Horizontal-Induced
${\nu}$ Shower thresholds are suppressed by a corresponding
factor. More abundant single event Air-Induced ${\nu}$  Shower
(Vertical or Horizontal)  are facing different Air volumes and
quite different visibility. It must be taken into account an
additional factor three (for the event rate) (because of three
light neutrino states) in the Air-Induced ${\nu}$ Shower arrival
flux respect to incoming $\nu_{\tau}$ (and $\bar{\nu_{\tau}}$ )
in UPTAUs and HORTAUs, making the Air target not totally a
negligible calorimeter.

There are also a sub-category  of $\nu_{\tau}$ - $\tau$ "double
 bang" due to a first horizontal UHE $\nu_{\tau}$ charged current interaction
 in air  nuclei (the first bang) that is lost from the EUSO view;
 their UHE  secondary $\tau$ fly and decay leading to a Second Air-Induced Horizontal Shower, within the EUSO
 disk area. These  horizontal "Double-Single $\tau$ Air Bang"  Showers
 (or if you like popular terminology, these Air-Earth Skimming neutrinos or just Air-HORTAU event)
  are produced within a very wide Terrestrial Crown Air Area whose radius is exceeding $\sim 600- 800$ km
 surrounding  the EUSO Area of view. However it is easy to show
 that they will just double the  Air-Induced ${\nu}$  Horizontal Shower
 rate due to one unique flavour. Therefore the total Air-Induced Horizontal Shower (for
 all $3$ flavours and the additional $\tau$ decay in flight) are summirized and considered
 in next figure. The relevant UHE neutrino signal, as discussed below, are due to the
 Horizontal Tau Air-Showers originated within the (much denser)Earth
 Crust:the called  HORTAUs (or Earth Skimming $\nu_{\tau}$).\\
\section{Horizontal Tau from Earth Skin: HORTAUs}
As already mention the UHE $\nu$ astronomy maybe greatly
amplified by $\nu_{\tau}$ appearance via flavour mixing and
oscillations. The consequent scattering of $\nu_{\tau}$ on the
Mountains or into the Earth Crust may lead to Horizontal Tau
Air-Showers :HORTAUs (or so called Earth Skimming Showers
\cite{Fargion2001a} \cite{Fargion2001b} \cite{Fargion 2000-2002}
\cite{Feng et al 2002}). Indeed UHE $\nu_{\tau}$ may skip below
the Earth and escape as $\tau$ and finally decay in flight,
within air atmosphere, as well as inside  the Area of  view of
EUSO, as shown in Fig. below.
\begin{figure}
\centering
\includegraphics[width=7cm]{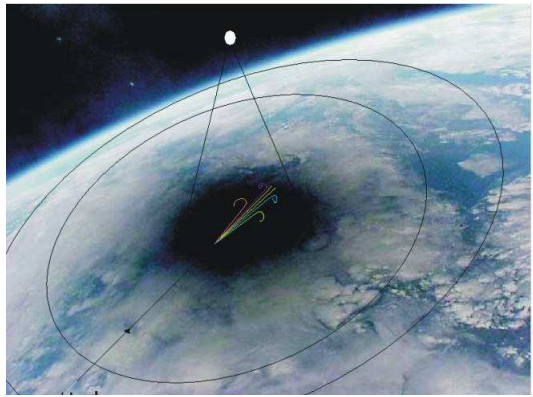}
\caption {A schematic Horizontal High Altitude Shower or similar
Horizontal Tau Air-Shower (HORTAUs) and its open fan-like jets
due to geo-magnetic bending seen from a high quota by EUSO
satellite. The image background is moon eclipse shadow observed
by Mir on Earth. The forked Shower is multi-finger containing a
inner $\gamma$ core and external fork spirals due to $e^+  e^-$
pairs (first opening) and  ${\mu}^+ {\mu}^-$ pairs.}
\label{fig:fig13}
\end{figure}
Any UHE-GZK Tau Air Shower induced event is approximately born
within a wide  ring  (whose radiuses extend between $R \geq 300$
and $R \leq 800$ km from the EUSO Area center). Because of the
wide area and deep $\tau$ penetration \cite{Fargion 2000-2002}
\cite{Fargion 2002b} \cite{Fargion 2002d} the amount of
interacting matter where UHE $\nu$ may lead to $\tau$ is huge
($\geq 2 \cdot 10^5$ $km^3$) ;however only a tiny fraction of
these HORTAUs will beam and Shower within the EUSO Area within
EUSO. We estimate  an effective Volumes for unitary area :
$$
\frac{V_{eff}}{A_{\oplus}}=
\int_0^{\frac{\pi}{2}}\frac{(2\,\pi\,\,R_{\oplus}\cos\theta)
\,l_{\tau}\,\sin{\theta}}{4\,\pi\,R^2_{\oplus}}\cdot e^{-
\frac{2\,R_{\oplus}\,\sin{\theta}}{L_{\nu_{\tau}}}}\,R_{\oplus}\,d\theta\,=
\frac{1}{2}\left({\frac{L_{\nu_{\tau}}}{2\,R_{\oplus}}}\right)
^2\,l_{\tau}\int_0^{\frac{2\,R_{\oplus}}{L_{\nu_{\tau}}}}t\cdot
e^{-\,t}d\,t
$$

Where $V_{eff}$ is the effective volume where Ultra High Energy
neutrino interact while hitting the Earth and lead to escaping
UHE Tau: this volume encompass a wide crown belt, due to the
cross-section of neutrino Earth skimming along a ring of variable
radius $(R_{\oplus}\cos\theta)$ and a corresponding skin crown of
variable depth $l_{\tau} \sin\theta $. $A_{\oplus}$ is the total
terrestrial area, $l_{\tau}$ is the tau interaction lenght,
$L_{\nu_{\tau}}$ is the Ultra High Energy Neutrino tau
interaction (charged current) in matter. The geometrical
quantities  are  defined in reference \cite{Fargion 2002b} while
$A _{EUSO}$ is the EUSO field of view Area. The more severe and
realistic suppression factors should be included in present
analytical derivation: first , the exponential decay of air
density at highest (derived in Appendix , see reference
\cite{Fargion 2000-2002}); secondly the Earth Crust opacity for
UHE neutrino tau at each integral step, already introduced in
previous exact $V_eff$ estimate; third, the neutrino energy
degradation by neutral current interactions and mainly the
charged current production of tau whose energy losses are
constraining the detectable volume from EUSO . Assuming two limit
cases (Earth all in water or only a thin water layer of
terrestrial water no deeper than $4.5 km$ in depth ) one finds:

\begin{eqnarray}
V_{eff-Max}= \frac{1}{2}\,A_{Euso}\,\left({1-e^{-
\frac{L_0}{c\,\tau_{\tau}\,\gamma_{\tau}}}}\right)
\,\left({\frac{L_{\nu_{\tau}}}{2\,R_{\oplus}}}\right)^2 \cdot
l_{\tau}\left[{1\,-\,e^{-
\frac{2\,R_{\oplus}}{L_{\nu_{\tau}}}}(1\,+\,\frac{2\,R_{\oplus}}{L_{\nu_{\tau}}})
}\right]
\end{eqnarray}

$$V_{eff-Min}= \frac{1}{2}\,A_{Euso}\,\left({1-e^{-
\frac{L_0}{c\,\tau_{\tau}\,\gamma_{\tau}}}}\right)
\,\left({\frac{L_{\nu_{\tau}}}{2\,R_{\oplus}}}\right)^2\cdot
l_{\tau}\left[{1\,-\,e^{- \frac{2\,R_{\oplus} sin(\theta_1)
}{L_{\nu_{\tau}}}}(1\,+\,\frac{2\,R_{\oplus}
sin(\theta_1)}{L_{\nu_{\tau}}}) }\right]$$
Where $\theta_1$ is the opening angle observed at horizon toward
the first terrestrial radius step (ocean -inner rock) and it is
$\theta_1 \cong 1.076 ^o$.
The  effective Volume assuming an outcoming Tau at energy
$E_{\tau}= 3 \cdot10^{19} eV$, becomes  $V_{eff}=5.5 \cdot 10^{2}
km^3$ and the event rate is  $N_{ev} = 7.7$ in three years.
 A more accurate estimate offers a slightly lower value .
 The maximum of the volume occurs at a Tau energy $E_{\tau}=2\cdot 10^{17} eV$
and it correspond to a volume $V_{eff}=7.5\cdot 10^{3} km^3$.
As it is manifest from the above curve the maximal event numbers
takes place at EeV energies. Therefore from here we derived the
primary interest for EUSO to seek lowest threshold (as low as
$10^{19}eV$). The above expression for the horizontal tau
air-shower contains , at lowest energies, the UPTAUs case. Indeed
it is possible to see that the same above  effective volume at
lowest energies simplify and  reduces to:
$ V_{eff}=\frac{1}{2}\,A_{Euso}\,\left({1-e^{-
\frac{L_0}{c\,\tau_{\tau}\,\gamma_{\tau}}}}\right) \,l_{\tau} $
Because one observes the Earth only from one side   the Area
factor in eq. $1$ should be $A_{\oplus} = {2\,\pi\,R^2_{\oplus}}$
and therefore the half in above formula may be dropped and the
final result is just the common expression $V_{eff} =
A_{Euso}\,l_{\tau}$.
\section{Conclusions: Event Rate for GZK Neutrinos at EUSO}
The above effective volume should be considered for any given
neutrino flux to estimate the outcoming EUSO event number. Here
we derive first the analytical formula. These general expression
will be plot assuming a minima GZK neutrino flux $ \phi_{\nu}$
just comparable to observed UHECR one $ \phi_{\nu}\simeq
\phi_{UHECR} \simeq 3\cdot 10^{-18} cm^{-2} s^{-1} sr{-1}$ at the
same energy ($E_{\nu}= E_{UHECR}\simeq 10^{19} eV$). This
assumption may changed at will (model dependent) but the event
number will scale linearly accordingly to the model. From here we
may estimate the event rate in EUSO by a simple extension:
\begin{eqnarray}
N_{eventi}\,=\,\Phi_{\nu}\,4\,\pi\,\eta_{Euso}\Delta
\,t\,\left({\frac{V_{eff}}{L_{\nu}}}\right)
\end{eqnarray}
Where $\eta_{Euso}$ is the  duty cycle fraction of EUSO,
$\eta_{Euso} \simeq 10\%$, $\Delta \,t\ \simeq 3$  $years$ and
$L_{\nu}$ has been defined in figure above.
$$
N_{eventi}=
\frac{1}{2}\,4\pi\,\eta_{Euso},{\Delta{t}}\Phi_{\nu_\tau}\,A_{Euso}\,\cdot
      \left({1-e^{-\frac{L_0}{c\,\tau_{\tau}\,\gamma_{\tau}}}}\right)\left({\frac{l_{\tau}}{L_{\nu_{\tau}}}}\right)\left({\frac{L_{\nu_{\tau}}}{2\,R_{\oplus}}}\right)^2\cdot \nonumber\\
\left[{1\,-\,e^{-\frac{2\,R_{\oplus}}{L_{\nu_{\tau}}}}(1\,+\,\frac{2\,R_{\oplus}}{L_{\nu_{\tau}}})}\right]
$$
It should be remind that all these event number curves for EUSO
are already suppressed by a factor $0.1$ due to minimal EUSO duty
cycle.
However the Air-Shower induced neutrino may reflect all three
light neutrino flavours, while HORTAUs are made only by
$\nu_{\tau}$,$\bar{\nu_{\tau}}$ flavour. Nevertheless the
dominant role of HORTAUs overcome (by a factor $\geq 6$) all other
Horizontal EUSO neutrino event at lowest energy edges $10^{19}$eV:
their expected event rate are, at $\Phi_{\nu}\geq 3 \cdot 10^{3}$
eV $cm^{-2} s^{-1}$ neutrino fluence, as in Z-Shower model,  a
few hundred event a year and they may already be comparable or
even may exceed the expected Horizontal CR rate. Dash curves for
both HORTAUs and Horizontal Cosmic Rays are drawn assuming an
EUSO threshold at $10^{19}$eV. Because the bounded $\tau$ flight
distance (due to the contained terrestrial atmosphere height) the
main signal is  better observable at $1.1 \cdot 10^{19}$eV than
higher energies as emphasized  at different  curves in figure
above.
  It should be noticed that  HORTAUs are very long  high altitude showers. Their lenght may exceed
  two hundred kilometers. This trace may be larger than the EUSO
  radius of Field of View. Therefore there may be both contained
  and partially contained events. There may be also crossing
  HORTAUs at the edges of EUSO  disk area. However most
  of the events will be partially contained, either just on their
  birth or at their end, equally balanced in number. Because of the HORTAU Jet forked shower,
  its up-going direction, its fan like structure (see Fig.\ref{fig:fig7}, Fig.\ref{fig:fig8}, Fig.\ref{fig:fig13}), these partially
  contained shower will be the manifest and mostly useful and
  clear event. The area of their origination, four times larger than
  EUSO field of view, will be mostly
  outside the same EUSO area. Their total number count will double the
  event rate $N_{ev}$ (and the corresponding $V_{eff}$) of
  HORTAUs. The additional crossing event will make additional events (a small fraction) of the effective
  volume of HORTAUs at $10^{19} eV$ the most rich neutrino signal few times larger the Air induced events.
  The same doubling will apply only to UHECR horizontal shower
  while the downward Ultra High Energy Neutrino will not share
  this phenomena (out of those $\simeq 6\%$ of the Horizontal Air Neutrino
  Shower).
\begin{figure}
\centering
\includegraphics[width=10cm]{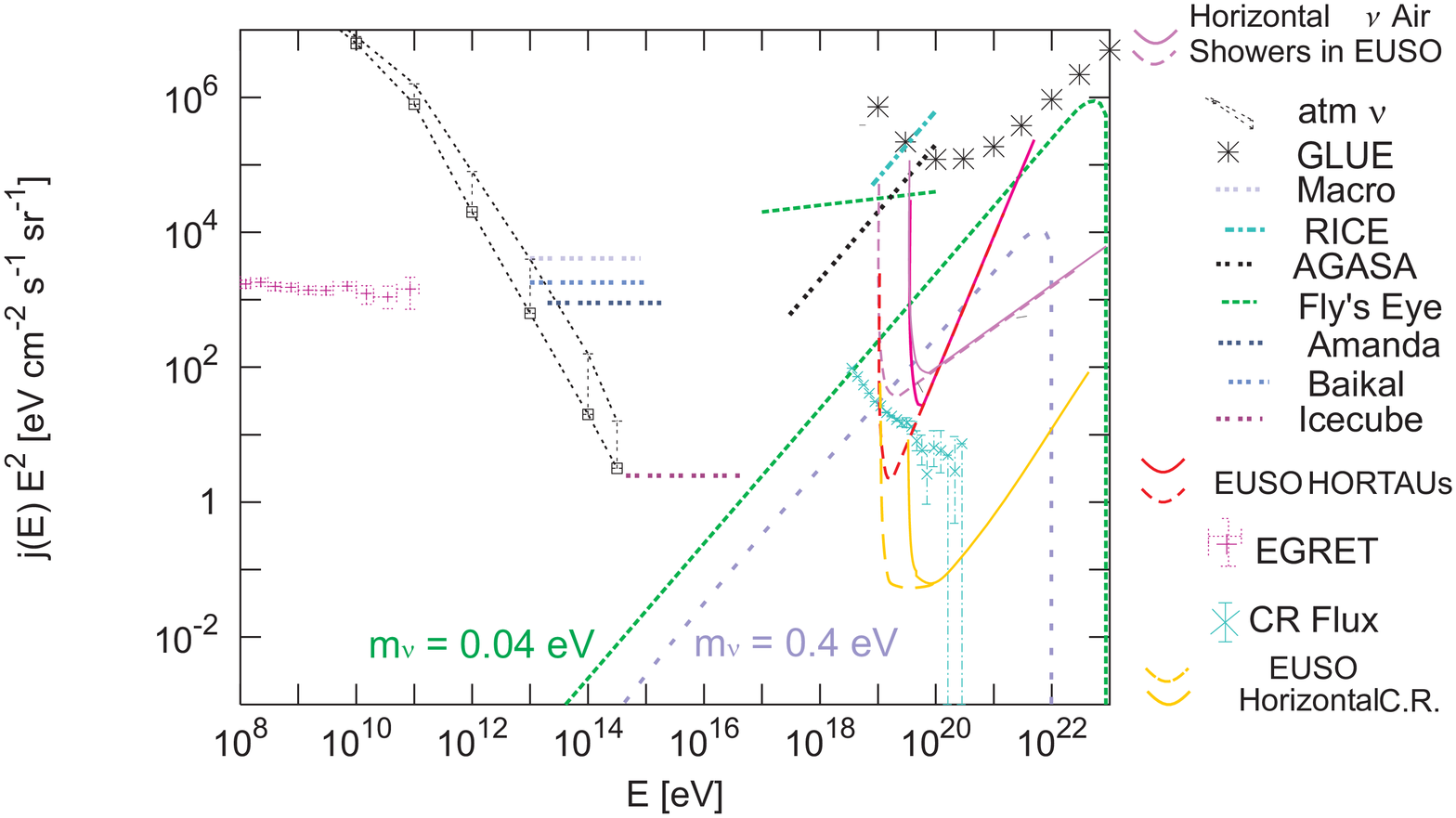}
\caption {EUSO thresholds for Horizontal Tau Air-Shower shower,
HORTAUs (or Earth Skimming Showers) over all other $\gamma$, $\nu$
and Cosmic Rays (C.R.) Fluence and bounds. The Fluence threshold
for EUSO has been estimated for a three year experiment lifetime.
Competitive experiment are also shown as well as the Z-Shower
expected spectra in light neutrino mass values ($m_{\nu} = 0.04,
0.4$ eV). As above dash curves for both HORTAUs and Horizontal
Cosmic Rays are drawn assuming an EUSO threshold at $10^{19}$eV.
}\label{fig:fig16}
\centering
\includegraphics[width=10cm]{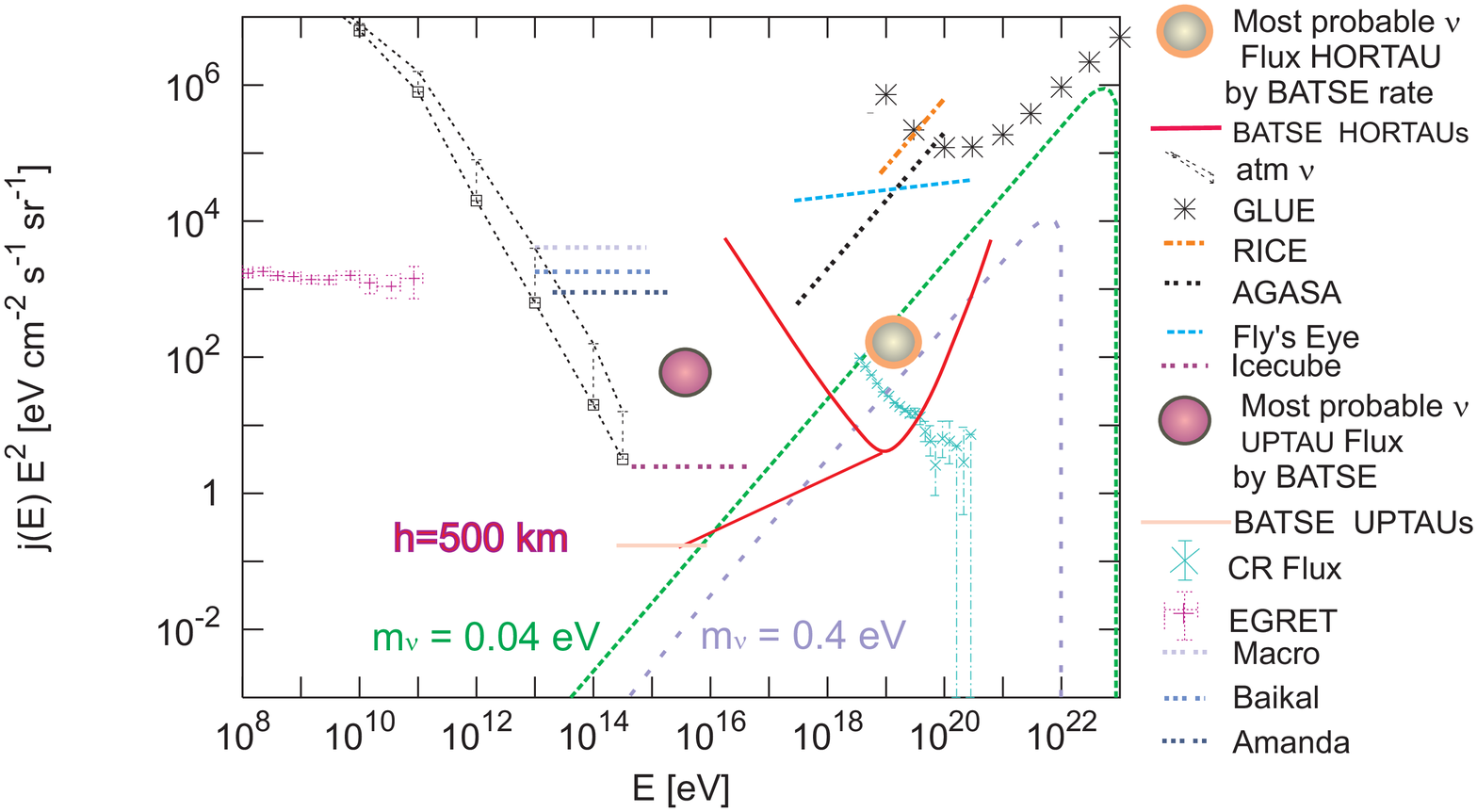}
\caption {Neutrino Flux derived by  BATSE Terrestrial Gamma
Flashes assuming them as $\gamma$ secondaries of upward Tau
Air-showers. These fluxes are estimated using the data from
Terrestrial Gamma Flash (1991-2000) normalized during  their most
active trigger and TGF hard activities. The UPTAUs and HORTAUs
rate are normalized assuming that the events at geo-center angle
above $50^o$ might be of HORTAU nature. } \label{fig:fig17}
\end{figure}
Even for the most conservative scenario where a minimal GZK-$\nu$
fluence must take place (at least at:$E_{\nu}\,\Phi_{\nu} \simeq
30 eV cm^{-2} s^{-1}sr^{-1}$, just comparable to well observed
Cosmic Ray fluence), half a dozen of such UHE astrophysical
neutrino must be observed  during three year of EUSO data
recording. These HORTAUs will be not be observable by other
competitive experiment as AUGER. Therefore to improve the HORTAU
visibility in EUSO one must: a) Improve the fast pattern
recognition of Horizontal Shower Tracks with their few distant
dots with forking signature, b) Enlarge the Telescope Radius to
embrace also lower $10^{19}$ eV energy thresholds where UHE
HORTAU neutrino signals are enhanced, c) Consider a detection
at  angular $\Delta\theta$ and at height $\Delta h$ level within
an accuracy $\Delta\theta \leq 0.2^o$, $\Delta h \leq 2$ km. Even
all  the above results have been derived carefully  following
\cite{Fargion 2002b} \cite{Fargion 2002c} \cite{Fargion 2002e} in
a minimal realistic framework they may be used within $20\%$
nominal value due to the present uncertain in  EUSO detection
capabilities. In conclusion  UHECR and Neutrino Astronomy face a
new birth. The Neutrino Astronomy may be widely discovered by
Upward and Horizontal $\tau$ Air-Showers. The Tau neutrinos, born
abundantly by flavour mixing will probe such Astronomy above PeVs
up to EeVs energies, where astrophysics rule over atmospheric
neutrino noise. The same UHE $\overline{\nu_{e}}$  at
$E_{\nu_{e}} = \frac{{M_W}^2}{2 \cdot m_e} \simeq 6.3 PeV $  must
be a peculiar neutrino astronomy born beyond Mountain Chains
\cite{Fargion et all 1999},\cite{Fargion 2000-2002} with its
distinctive signature. Past detectors as GRO BATSE experiment
might already found some direct signals of such rare UPTAUs or
HORTAUs; indeed their observed Terrestrial Gamma Flash event rate
translated into a neutrino induced  upward shower (see
Fig.\ref{fig:fig17}) leads to a most probable flux both at PeVs
energies  just at a level comparable to most recent AMANDA
threshold sensitivity: for horizontal TGF events at $10^{19}$ eV
windows, the signals fit the Z-Burst model needed fluence (for
neutrino  at $0.04-0.4$ eV masses).It is remarkable to note that
most of the celestial sources associate to TGF are on the
galactic plane and in particular toward the galactic center
(namely black hole candidate source $GX 339-4$ and most active
known galactic sources \cite{Fargion 2003a}). Future EUSO
telescope detector, if little enlarged will easily probe even the
smallest, but necessary Neutrino GZK fluxes with clear sensitivity
(seeFig.\ref{fig:fig16}). We therefore expect that a serial of
experiment will soon turn toward this last and neglected, but
most promising Highest Energy Tau Neutrino Astronomy searching
for GZK or Z-Showers neutrino signatures.
\\
\\
\\

%
\bibliography{xbib}
\end{document}